\documentclass[conference]{IEEEtran}
\IEEEoverridecommandlockouts
\usepackage{float}
\usepackage{cite}
\usepackage{amsmath,amssymb,amsfonts}
\usepackage{graphicx}
\usepackage{algorithmic}
\usepackage{textcomp}
\usepackage{xcolor}
\usepackage{orcidlink}
\usepackage[linesnumbered,ruled,vlined] {algorithm2e}
\begin{document}
\pagestyle{plain}

\graphicspath{{./images/}}

\title{Secure Distributed Learning for CAVs: Defending Against Gradient Leakage with Leveled Homomorphic Encryption}

\author{
    Muhammad Ali Najjar\IEEEauthorrefmark{1}\orcidlink{0009-0008-3800-306X},
    Ren-Yi Huang\IEEEauthorrefmark{2}\orcidlink{0009-0006-8341-5835}, 
    Dumindu Samaraweera, \textit{Member, IEEE}\IEEEauthorrefmark{1}\orcidlink{0000-0003-4097-5585},
    and Prashant Shekhar\IEEEauthorrefmark{1}\orcidlink{0000-0003-2353-6740}

    \thanks{\IEEEauthorrefmark{1}Department of Mathematics, Embry-Riddle Aeronautical University, Daytona Beach, FL (emails: najjarm@my.erau.edu, samarawg@erau.edu, shekharp@erau.edu).}
    \thanks{\IEEEauthorrefmark{2}Department of Electrical Engineering, University of South Florida, Tampa, FL (email: hr219@usf.edu).}
}

\maketitle
\footnotetext{Code and data of this work available at: \url{https://github.com/Rahn80643/Federated-Learning-PyTorch-HE-Smap}}

\begin{abstract}
Federated Learning (FL) enables collaborative model training across distributed clients without sharing raw data, making it a promising approach for privacy-preserving machine learning in domains like Connected and Autonomous Vehicles (CAVs). However, recent studies have shown that exchanged model gradients remain susceptible to inference attacks such as Deep Leakage from Gradients (DLG), which can reconstruct private training data. While existing defenses like Differential Privacy (DP) and Secure Multi-Party Computation (SMPC) offer protection, they often compromise model accuracy. To that end, Homomorphic Encryption (HE) offers a promising alternative by enabling lossless computation directly on encrypted data, thereby preserving both privacy and model utility. However, HE introduces significant computational and communication overhead, which can hinder its practical adoption. To address this, we systematically evaluate various leveled HE schemes to identify the most suitable for FL in resource-constrained environments due to its ability to support fixed-depth computations without requiring costly bootstrapping. Our contributions in this paper include a comprehensive evaluation of HE schemes for real-world FL applications, a selective encryption strategy that targets only the most sensitive gradients to minimize computational overhead, and the development of a full HE-based FL pipeline that effectively mitigates DLG attacks while preserving model accuracy. We open-source our implementation to encourage reproducibility and facilitate adoption in safety-critical domains.

\end{abstract}

\begin{IEEEkeywords}
Federated Learning, Homomorphic Encryption, Gradient Leakage Attack, Privacy, CAVs
\end{IEEEkeywords}

\section{Introduction}
Federated Learning (FL) has emerged as a promising paradigm to address data privacy concerns in Machine Learning (ML) by enabling distributed and collaborative model training without the need of direct data sharing. Instead of centralizing data, FL distributes the learning process across participating clients, where each client trains a local model on its data, and only model updates are shared with the central server for aggregation. This approach inherently strengthens data privacy, making it widely applicable in practical domains such as Connected and Autonomous Vehicles (CAVs), Unmanned Aerial Vehicles (UAVs), and similar systems. However, the exchange of model parameters in FL is not without risk, as these parameters themselves can be targets for inference or reconstruction attacks and may still leak sensitive information. For instance, recent studies reveal significant privacy vulnerabilities, particularly through gradient leakage attacks, such as Deep Leakage from Gradients (DLG), where adversaries exploit exchanged model gradients to reconstruct private training data \cite{huang2024exploring, zhu2019deep}.

To address the growing security and privacy concerns in FL, several mitigation strategies have been actively explored. Differential Privacy (DP) is one of the most widely used and straightforward techniques, where carefully calibrated noise is added to model updates or gradients to prevent the leakage of individual data points. While DP offers strong theoretical guarantees, it may introduce a trade-off between privacy and model accuracy. Secure Multi-Party Computation (SMPC) enables multiple parties to collaboratively compute a function over their inputs while keeping those inputs private, ensuring that no individual party learns more than necessary. Homomorphic Encryption (HE), a cryptographic technique first introduced by Rivest et al. in 1978 \cite{rivest1978method}, takes this a step further by allowing computations to be performed directly on encrypted data, enabling a server to aggregate encrypted model updates without ever accessing the raw data. Each of these methods offers different levels of protection and computational overhead, and ongoing research focuses on combining them to strike a balance between privacy, efficiency, and model utility in practical FL deployments. However, compared to other methods, HE is a lossless mechanism that preserves the full utility of the data while providing strong privacy guarantees, making it particularly suitable for a wide range of practical applications.

Different HE schemes exhibit varying performance characteristics, and selecting the most suitable scheme and applying optimizations are crucial to making HE-based federated learning viable in resource-constrained environments such as CAVs. HE has undergone significant evolution over the past two decades, with each generation addressing critical challenges related to functionality, efficiency, and practical deployment. These HE schemes can broadly be classified into three main generations, each representing a significant advancement in capability and practicality:

\begin{itemize}
    \item First Generation – Partially Homomorphic Encryption (PHE):
    The earliest form of HE supported only a single type of mathematical operation (either addition or multiplication) an unlimited number of times. Examples include RSA (multiplicative homomorphism \cite{rivest1978method}) and Paillier (additive homomorphism \cite{paillier1999public}). Although conceptually useful, these schemes were limited in real-world applicability because of their inability to handle general computations.
    
    \item Second Generation – Somewhat and Fully Homomorphic Encryption (SHE \& FHE):
    This generation introduced the ability to perform both addition and multiplication on encrypted data, enabling more expressive computations. Somewhat Homomorphic Encryption (SHE) supported limited-depth computations before ciphertexts became too noisy. The major breakthrough came in 2009, when Craig Gentry introduced the first Fully Homomorphic Encryption (FHE) scheme  \cite{gentry2009fully}. His construction, based on ideal lattices and a revolutionary method called bootstrapping, allowed for arbitrary-depth computation on encrypted data and thereby laying the foundation for general-purpose privacy-preserving computation.
    
    \item Third Generation – Practical/Optimized FHE:
    More recent advances have focused on improving the efficiency, scalability, and usability of FHE. This includes optimizations such as SIMD-style parallel processing \cite{brakerski2014leveled}, batching techniques, and support for approximate arithmetic (e.g., Cheon-Kim-Kim-Song (CKKS) scheme for encrypted floating-point operations \cite{cheon2017homomorphic}). These developments have significantly improved performance and made HE increasingly viable for practical applications such as privacy-preserving machine learning, secure cloud computing, and federated learning.
\end{itemize}

Although promising, integrating HE into FL architectures remains highly challenging due to significant computational and communication overheads. To partially mitigate these challenges, Leveled Homomorphic Encryption (LHE) can be employed. LHE is a subclass of HE that supports both addition and multiplication of encrypted data, but only up to a fixed number of operations, known as multiplicative depth. Unlike FHE, which allows unlimited operations via the expensive bootstrapping process, LHE eliminates bootstrapping entirely, offering improved efficiency and lower latency. This makes LHE especially suitable for practical scenarios where the computation depth is known in advance, such as privacy-preserving machine learning. By carefully tuning the encryption parameters to match the task complexity, LHE offers a practical trade-off between security and performance for FL systems.

The key contributions of this paper are as follows:
\begin{enumerate}
    \item We conduct an extensive evaluation of practical HE schemes, focusing on their functionality, efficiency, and suitability for real-world deployment, with Connected and Autonomous Vehicles as a primary use case.
    \item We propose and implement a selective encryption strategy that prioritizes the most sensitive gradients, effectively reducing the computational overhead of HE while preserving model utility.
    \item We design and implement an end-to-end privacy-preserving pipeline that effectively mitigates Deep Leakage from Gradients (DLG) attacks using HE. Our solution maintains near-identical model accuracy while providing strong privacy guarantees, and we make the implementation publicly available to foster reproducibility and adoption.
\end{enumerate}

The rest of the paper is organized as follows. Section II presents essential background on federated learning, homomorphic encryption, and related work in this domain. Section III reviews adversarial attacks targeting FL architectures, with a particular emphasis on applications in CAVs. Section IV introduces our proposed selective parameter encryption strategy using the CKKS scheme, along with the implementation framework. Section V details the experimental setup, while Section VI discusses the experimental results and key findings. Finally, Section VII concludes the paper and outlines directions for future research.

\section{Background and Related Work}
This section provides an overview of the background and related work, with a particular focus on the attack landscape in federated learning, especially as it pertains to CAV applications.

\begin{figure}[H]
    \centering
    \includegraphics[width=1.0\linewidth]{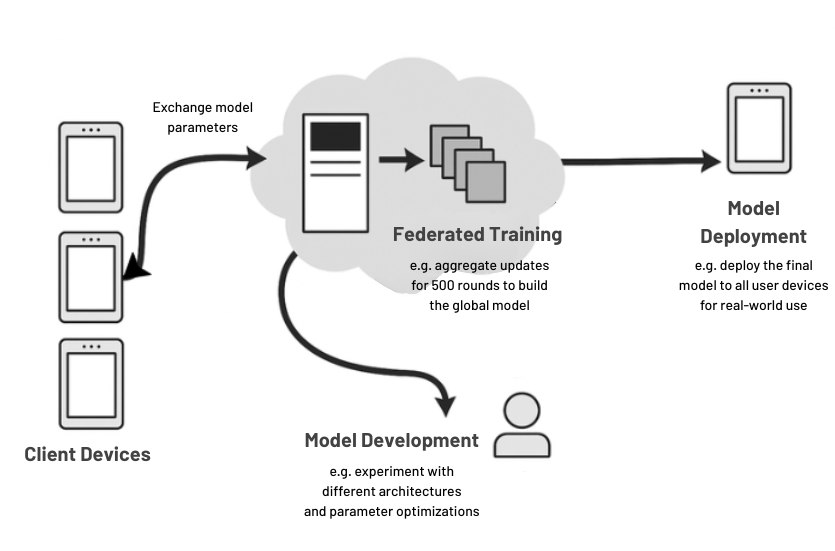}
    \caption{An overview of the federated learning workflow in distributed systems, including its decentralized training architecture and collaborative model aggregation process.}
\end{figure}

\subsection{Federated Learning}
FL trains a global machine learning model across decentralized nodes or client devices, each with private datasets. Rather than transferring raw data to a central server, FL aggregates local model updates or gradients to enhance privacy. Over the years, various aggregation algorithms have been proposed to strike a balance between collaboration efficiency and data privacy in FL. A foundational method, Federated Averaging (FedAvg), introduced by McMahan et al. \cite{mcmahan2017communication}, optimizes local Stochastic Gradient Descent (SGD) by aggregating model updates while discarding those from straggler devices, thereby improving communication efficiency. To address challenges arising from statistical and system heterogeneity among clients, FedProx extends FedAvg by incorporating a proximal term that regularizes local updates \cite{li2020federated}. Further advancements, such as SCAFFOLD, introduce control variates to mitigate client drift and enhance convergence, especially under non-IID data distributions \cite{karimireddy2020scaffold}.

However, regardless of these aggregation methods, FL's privacy vulnerabilities come primarily from the exposure of gradients, which allows attackers to infer sensitive attributes from model updates, underscoring the critical need for robust privacy preservation methods.

\subsection{Encrypting Model Parameters with HE}
In the context of FL, homomorphic encryption enables computations directly on encrypted data without requiring decryption of intermediate values, thereby significantly enhancing privacy guarantees. As previously mentioned, HE schemes are generally categorized into three types: Partially Homomorphic Encryption (PHE), which supports only a single type of operation (either addition or multiplication); Somewhat Homomorphic Encryption (SHE), which allows a limited number of both operations; and Fully Homomorphic Encryption (FHE), which supports arbitrary computations on encrypted data without limitation \cite{acar2018survey}. Classical PHE schemes, such as RSA (multiplicative) and ElGamal (additive under certain settings), offer efficient encryption mechanisms but lack the operational flexibility required for complex machine learning tasks.

\begin{figure}[ht]
    \centering
    \includegraphics[width=1.0\linewidth]{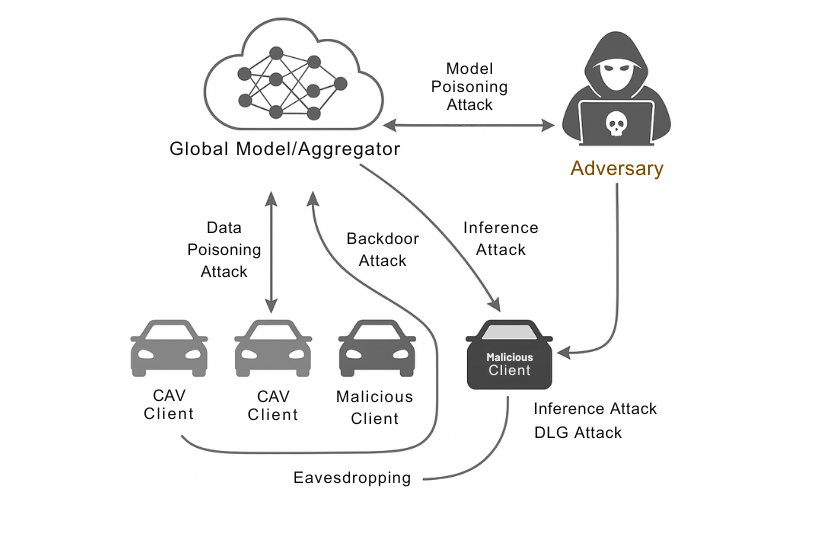}
    \caption{How federated aggregation operates at the server when model parameters are encrypted using homomorphic encryption.}
    \label{fig:HE}
\end{figure}

The introduction of FHE transformed encrypted computation by allowing any arbitrary arithmetic in ciphertext, although initially impractical due to computational overhead \cite{gentry2009fully}. Subsequent advances led to leveled HE schemes like Brakerski-Gentry-Vaikuntanathan (BGV) \cite{brakerski2012leveled}, Brakerski/Fan-Vercauteren (BFV) \cite{fan2012somewhat}, and particularly CKKS \cite{cheon2017homomorphic}, significantly optimizing computational efficiency for approximate arithmetic common in machine learning tasks. The CKKS scheme, a lattice-based leveled FHE scheme, supports approximate arithmetic on real numbers, making it highly suitable for machine learning workloads. Unlike schemes like BFV that are optimized for exact arithmetic, CKKS allows SIMD-style\footnote{Single Instruction, Multiple Data.} operations through efficient polynomial encoding, enabling the batching of encrypted gradients and reducing communication overheads during secure federated training. This means that multiple plaintext values can be packed into a single ciphertext and processed in parallel, significantly improving computational efficiency. Unlike BFV and BGV, which are designed for exact computations on integers, CKKS allows efficient approximate operations on floating-point numbers, making it well-suited for machine learning tasks that tolerate small numerical errors. The ability to encode and process vectors of real numbers directly gives CKKS a performance advantage in scenarios like federated learning, where large-scale, parallelizable computations are common.

\begin{figure}[H]
    \centering
    \includegraphics[width=1.0\linewidth]{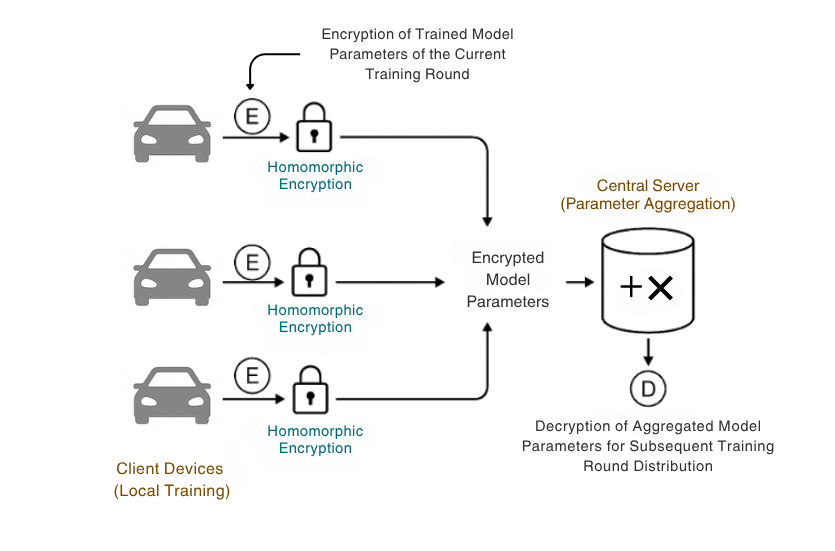}
    \caption{Security challenges in federated learning for distributed connected and autonomous vehicle environments.}
    \label{fig:CAV}
\end{figure}

While HE-based techniques such as CKKS offer a strong privacy guarantee by operating directly on encrypted data, they incur high computational costs-especially during bootstrapping, which remains a performance bottleneck. Recent profiling studies confirm that bootstrapping accounts for up to 50\% of runtime in practical deployments \cite{de2021does}, yet its cost is often justified by removing trust assumptions on the server. Moreover, arithmetic intensity analyses of bootstrapped CKKS circuits reveal that modern compute platforms like GPUs and FPGAs are often memory-bound due to limited cache capacity and large ciphertext sizes. These insights are driving research toward more feasible and practical implementations, where we employ a selective parameter encryption strategy—detailed later in Section IV of this paper.

\subsection{Integration of HE with Federated Learning for Practical Applications}
Although homomorphic encryption introduces inherent communication and computational overheads, its lossless utility has made it an attractive option for secure computation. As a result, several studies have specifically investigated the integration of HE within Federated Learning frameworks to enhance privacy without compromising model performance. Li et al. \cite{li2020federated} demonstrated the feasibility of employing HE schemes in federated settings, emphasizing that appropriate parameter selection and efficient encoding significantly affect overall system performance. Moreover, Zhang et al. \cite{zhang2020batchcrypt} introduced BatchCrypt, an optimized approach leveraging batching techniques to significantly reduce computational overhead in HE-FL systems, reinforcing the practicality of encrypted computations. Another notable contribution by Fang et al. \cite{fang2021privacy} highlights hybrid models integrating secure multi-party computation and HE to balance privacy guarantees and computational efficiency. Combining HE with FL addresses privacy concerns by securely encrypting gradient updates, mitigating leakage risks. Usually, such integration typically involves trade-offs between model accuracy, computational complexity, and communication overhead, requiring careful optimization to maintain system performance. Recent practical frameworks such as FedML-HE effectively manage these trade-offs, demonstrating real-world applicability of encrypted FL \cite{jin2023fedml}. These approaches selectively encrypt critical parameters, effectively balancing security and computational efficiency.


While homomorphic encryption offers compelling security benefits for federated learning in applications like CAVs, practical deployment requires more careful consideration of computational overhead and communication efficiency. In particular, CAVs typically generate enormous amounts of sensitive data from sensors, cameras, and vehicle-to-everything (V2X) communications. Thus, centralized training methods pose major risks regarding data privacy, bandwidth limitations, and latency \cite{sun2021survey}. Therefore, incorporating HE into FL for CAVs is essential, not only for protecting sensitive data such as user location, driving behavior, and sensor inputs during model updates, but also for ensuring compliance with privacy regulations like GDPR and industry-specific mandates. Moreover, it enables secure collaboration among multiple stakeholders, including car manufacturers, without compromising proprietary information. In addition, it supports the development of secure, real-time decision-making models by safeguarding communication payloads throughout the learning process \cite{alsamiri2023federated}.

\section{Adversarial Attacks on FL Architectures in CAV Applications}
Unlike centralized learning models, FL enables decentralized training across multiple clients, which, while enhancing data privacy, introduces a range of unique security vulnerabilities. One of the most prominent threats is the model poisoning attack \cite{bhagoji2019analyzing}, where a malicious client injects carefully crafted updates during training to manipulate the global model’s decision boundary, often without significantly degrading its performance on clean, non-targeted data. Closely related is the data poisoning attack \cite{fung2018mitigating}, in which the attacker alters, inserts, or removes local training data to degrade the model’s performance or introduce targeted bias, aiming to mislead the learning process by corrupting the underlying patterns.

Another concerning threat is the free-rider attack \cite{fang2020local}, where a client participates in the training rounds without performing meaningful computation, sending random, reused, or empty updates, yet still benefits from the improved global model trained by honest participants. Meanwhile, backdoor attacks embed hidden triggers into the model, causing misclassification only when specific patterns are present, without significantly altering the model's accuracy on standard inputs. In addition, inference attacks such as membership inference \cite{shokri2017membership} and gradient leakage aim to extract sensitive information from shared updates. Here, an adversary, either a participating client or the central server, analyzes model gradients or parameters to reconstruct private data, such as images or sensor readings, from other clients.

\begin{figure}[H]
    \centering
    \includegraphics[width=1\linewidth]{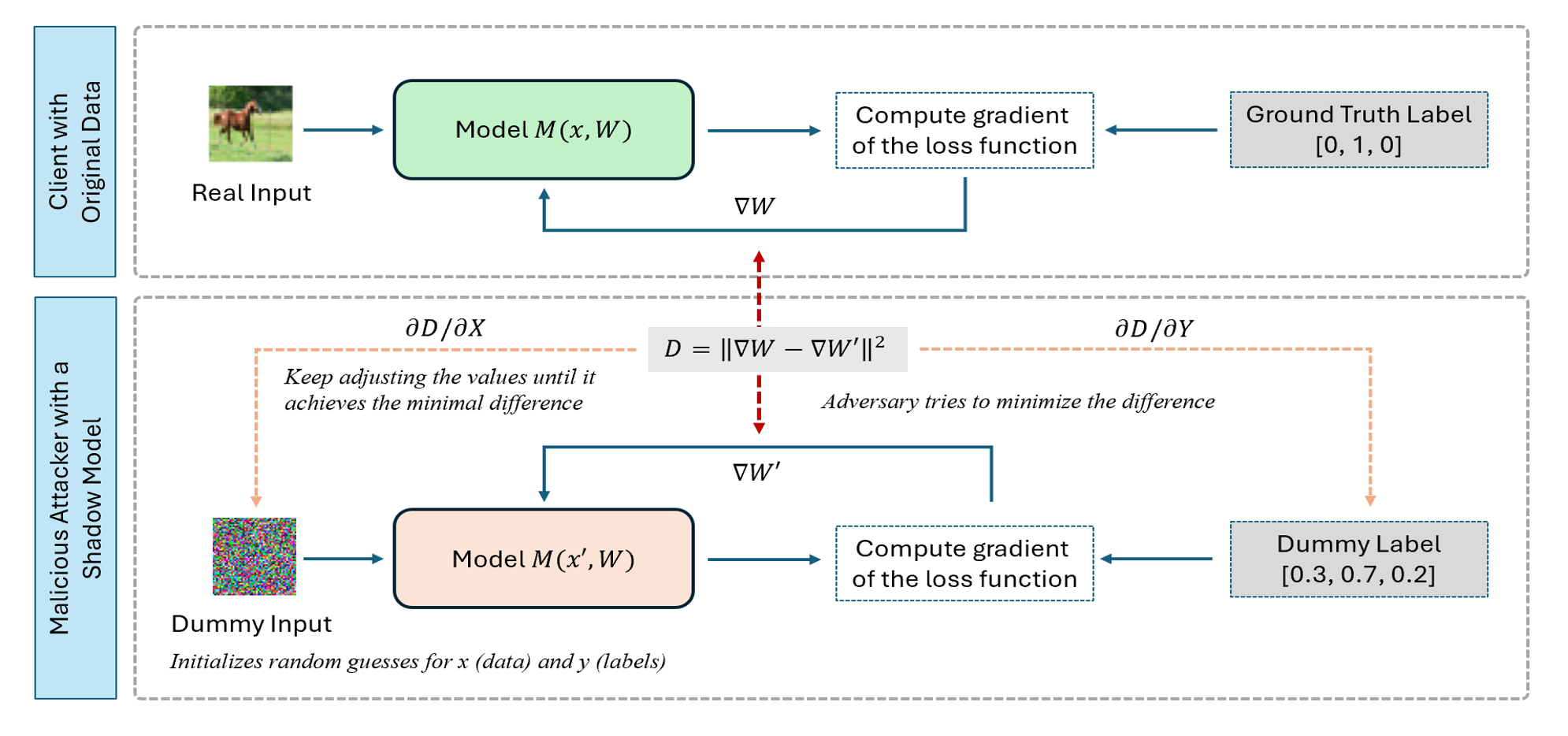}
    \caption{An overview of adversarial reconstruction of sensitive input data from shared gradients, beginning with random input initialization.}    
    \label{DLG}
\end{figure}

As illustrated in Fig. \ref{DLG}, Deep Leakage from Gradients (DLG) is an attack in which adversaries reconstruct sensitive input data from shared gradient information \cite{huang2024exploring}. Figure shows how DLG exploits the fact that gradients carry implicit information about training data, allowing adversaries to iteratively optimize dummy data to match observed gradients \cite{zhu2019deep}. Follow-up studies by Zhao et al. \cite{zhao2020idlg} improved the efficiency and accuracy of reconstruction attacks, emphasizing the need for robust countermeasures. Given its practical feasibility and significant threat to client privacy, this attack vector is the central focus of this study.

While federated learning is designed to enhance data privacy by keeping raw data local, its decentralized architecture limits centralized oversight, making the detection and mitigation of advanced attacks like DLG particularly challenging. As FL continues to be deployed in privacy sensitive domains, such as connected autonomous vehicles, developing robust countermeasures against gradient leakage attacks is essential to maintain trust in collaborative learning systems.

Moreover, while this paper primarily focuses on mitigating adversarial attacks in federated learning using homomorphic encryption, it is important to consider the broader adversarial threat landscape targeting CAVs. These attacks extend beyond FL models and can compromise various subsystems within CAVs. For instance, Qayyum et al. \cite{qayyum2020securing} outlined ML-related vulnerabilities in vehicular networks, while Chattopadhyay et al. \cite{chattopadhyay2020autonomous} emphasized the need for a security-by-design approach to AV systems. Sharma et al. \cite{sharma2019attacks} demonstrated how adversarial examples can bypass existing ML-based misbehavior detectors in CAVs. Adversarial attacks may target perception systems by manipulating inputs like camera images or LiDAR data, leading to misclassification of road signs or phantom object detection. Planning and control systems can also be attacked through false communication messages, causing unsafe maneuvers. V2X communication is particularly vulnerable to spoofing, where attackers impersonate legitimate sources to inject misleading data.

Securing CAVs against adversarial attacks demands a holistic approach, strengthening perception, control, and communication layers while embedding privacy-preserving methods like FL and HE for secure model training and data sharing. In this study, we focus on mitigating DLG attacks, as they pose a direct risk to data confidentiality during the gradient exchange phase, a critical component in FL workflows. By encrypting model updates using leveled HE through TenSEAL, we ensure that adversaries, including potentially honest but prudent servers, cannot access meaningful information from intercepted gradients.

\section{Selective Parameter Encryption with CKKS}
This paper investigates the challenges and practical optimizations of integrating homomorphic encryption into federated learning architectures, with a particular focus on connected and autonomous vehicle applications. This section outlines the implementation details and methodology used in our experimental evaluations, which are discussed in subsequent sections.

Over the years, several HE libraries have been developed to support encrypted federated learning, each offering unique strengths. Microsoft SEAL \cite{sealcrypto}, a widely used C++ library, supports BFV and CKKS schemes and provides robust encryption operations, though it requires manual parameter tuning, which can be challenging for users unfamiliar with HE. IBM’s HElib \cite{halevi2020design} implements BGV and CKKS and includes advanced features like bootstrapping, but its steep learning curve and integration complexity limit accessibility. PALISADE \cite{PALISADE}, a general-purpose lattice cryptography library, supports multiple schemes (BFV, BGV, CKKS) and offers broad functionality, though it is less optimized for ease of use. In contrast, TenSEAL \cite{benaissa2021tenseal}, built on Microsoft SEAL and designed specifically for machine learning, offers a simplified Python interface, automated parameter selection, and efficient tensor operations, with seamless PyTorch integration. Among these, TenSEAL stands out as the most user-friendly and practical choice for implementing encrypted federated learning.

\subsection{Overview of the Workflow and Parameter Integration}
The proposed methodology integrates the CKKS leveled HE scheme within an FL framework using the TenSEAL library. TenSEAL, built on top of Microsoft SEAL, simplifies the handling of encrypted tensors, automates parameter management, and efficiently executes critical operations for machine learning models in privacy-preserving environments. Our experimental setup aligns with existing literature showing that ciphertext sizes grow significantly with increasing multiplicative depth. This requires frequent bootstrapping or circuit redesign to stay within a feasible computational envelope \cite{agrawal2023architecting}.

The CKKS algorithm encompasses several key operations, including the encoding of real numbers into polynomial representations, key generation and switching, noise management through rescaling, and computational bootstrapping to extend ciphertext usability. CKKS is especially well-suited for approximate arithmetic on floating-point numbers, enabling efficient computation of operations such as inner products and polynomial approximations (e.g., sigmoid functions). Achieving an effective balance between security, precision, and computational efficiency requires careful tuning of encryption parameters, such as the polynomial modulus degree, ciphertext modulus, and scaling factors. In this work, CKKS parameters were integrated using the TenSEAL library, configured as follows:

\begin{itemize}
    \item \textbf{Polynomial modulus degree (N)}: Set to 8192 based on the required 128-bit security level and TenSEAL limitations. Higher N values provide greater security but increase memory and computational demands.
    
    \item \textbf{Ciphertext modulus (Q)}: Selected based on the anticipated multiplicative depth of operations. It ensures that the encrypted computations maintain accuracy over multiple operations. It is constructed as a product of prime moduli, whose bit-lengths sum to $Q_{\text{bits}} = 60 + 52 + 60 = 172$ bits by default. This parameter was set to 172 in all experimental analysis.
    
    \item \textbf{Scaling factor}: Determines the precision of floating-point operations. A higher scaling factor results in better numerical stability but requires larger ciphertext sizes. CKKS encodes real numbers by scaling them to integers with a multiplicative factor $\Delta = 2^{52}$ (default). This scaling factor balances the precision and the available noise budget. Scaling factor  was set to 52 in all our experimental analysis.

\end{itemize}

\subsection{Implementation with Selective Encryption}
To reduce the additional computational and communication overhead introduced by homomorphic encryption, we adopt a selective encryption strategy that targets only the most critical model parameters, rather than encrypting the entire parameter set. This strategy can be implemented using two approaches:

\begin{itemize}
    \item \textbf{Jacobian-based Sensitivity Maps}: Gradients of model outputs with respect to each parameter are computed across training batches. Parameter-wise second-order derivatives are then averaged to rank parameters based on their sensitivity/importance.

    \item \textbf{Magnitude-based Sensitivity Maps}: After each mini-batch, layer-wise parameter magnitudes are evaluated to identify influential weights, without requiring the computation of second-order gradients.
\end{itemize}

These sensitivity maps enable partial encryption, where only the top-ranked parameters are encrypted. This significantly reduces computational costs while maintaining model utility and privacy protection. Building on our previous findings, this work adopts the magnitude-based sensitivity approach due to its lower computational overhead and ease of implementation compared to gradient-based alternatives. \cite{molchanov2019importance}

Algorithm~\ref{alg:selective_encryption} provides an overview of the selective CKKS encryption process, which comprises the following key components.

\begin{enumerate}
    \item \textbf{Client Initialization}: Clients generate local models based on private data, initializing HE parameters, including public/private key pairs, using TenSEAL with specific polynomial modulus degree and scaling factors.
    
    \item \textbf{Encrypted Local Training}: Each client performs local training rounds, encrypting either full set of gradients or selectively encrypted parameters based on the generated sensitivity maps. The encryption leverages CKKS for approximate arithmetic, balancing precision and computational efficiency.
    
    \item \textbf{Encrypted Aggregation}: Clients securely transmit encrypted updates to the central aggregator. Aggregation operations such as addition are performed homomorphically without decryption, ensuring data confidentiality throughout training.
    
    \item \textbf{Decryption and Update}: After aggregation, the server decrypts the combined model updates using the secret key and updates the global model, which is then broadcasted to the clients for the next round of training.
\end{enumerate}

\begin{algorithm}
\SetAlgoLined
\KwIn{Global model $\mathcal{M}_0$, client datasets $\{\mathcal{D}_i\}_{i=1}^{N}$, total rounds $T$, encryption ratio $r$}
\KwOut{Trained global model $\mathcal{M}_T$}

Initialize encryption context with CKKS parameters (ring dimension, scale, etc.)\;
Generate public/private keys $(pk_i, sk_i)$ for each client $i$\;
Broadcast $\mathcal{M}_0$ and $pk_i$ to all clients\;

\For{each round $t = 1$ \KwTo $T$}{
    \For{each client $i \in \{1, \dots, N\}$ in parallel}{
        Receive $\mathcal{M}_{t-1}$ from server\;
        Train local model $\mathcal{M}_i^{(t)}$ on data $\mathcal{D}_i$\;
        Compute local gradient $\nabla \mathcal{M}_i^{(t)}$\;
        
        Generate sensitivity map $S_i$ using gradient magnitude or Jacobian\;
        Identify top-$r$\% sensitive elements in $\nabla \mathcal{M}_i^{(t)}$\;

        Encrypt sensitive elements with CKKS using $pk_i$\;
        Send \texttt{(Encrypted, Plaintext)} gradient tuple to server\;
    }

    \textbf{Server side:} \\
    Aggregate encrypted gradients homomorphically\;
    Aggregate unencrypted gradients normally\;
    Decrypt aggregate encrypted portion using $sk_i$ or shared mechanism\;
    Combine both to compute full aggregated gradient $\nabla \mathcal{M}_t$\;
    Update global model $\mathcal{M}_t \gets \mathcal{M}_{t-1} - \eta \nabla \mathcal{M}_t$\;
    Broadcast updated $\mathcal{M}_t$ to all clients\;
}
\Return{$\mathcal{M}_T$}\;
\caption{Federated Learning with Selective CKKS Encryption}
\label{alg:selective_encryption}
\end{algorithm}

Our implementation framework incorporates several additional key features to ensure modularity, flexibility, and detailed performance analysis. Keypair generation is managed based on the selected HE library (TenSEAL) \cite{ibarrondo2021pyfhel}, enabling seamless integration and adaptability. To thoroughly evaluate overhead, we individually record encryption and decryption times, local training durations, and HE aggregation times. Partial encryption is applied using sensitivity maps, which can be generated via either magnitude-based or Jacobian-based methods, allowing selective protection of critical model parameters. Additionally, model encryption is performed layer-wise, supported by modular serialization functions that facilitate checkpointing and resumption. The complete source code and data for this work are available at \url{https://github.com/Rahn80643/Federated-Learning-PyTorch-HE-Smap}. Together, these design choices enable scalable, privacy-preserving federated learning with flexible encryption granularity and efficient encrypted aggregation, making the framework well-suited for practical applications such as connected and autonomous vehicles.

\section{Experimental Setup}
Our experimental setup leverages a high-performance computing platform configured with PyTorch and Python on an Ubuntu 20.04 server. The system is powered by Intel Xeon CPUs, 128 GB of RAM, and an NVIDIA RTX A6000 GPU, providing the computational capacity necessary for both deep learning and encrypted operations. For cryptographic computations, we employ the TenSEAL library, specifically utilizing the CKKS homomorphic encryption scheme. The encryption is configured with a polynomial ring dimension of 8192 and a scaling bit size of 52, offering a balance between computational efficiency and numerical precision.

To rigorously evaluate the proposed privacy-preserving federated learning framework, we adopt the CIFAR-10 benchmark dataset, which consists of 60,000 color images evenly distributed across 10 distinct classes. We simulate two types of data distribution scenarios to reflect realistic federated learning environments: (1) an IID setting where data is randomly and evenly distributed among clients, and (2) a Non-IID setting where class distributions are deliberately skewed to reflect heterogeneity. For this study, we focus on the IID configuration. The models used include EfficientNetB0, MobileNetV1, MobileNetV2, and ResNet34, with approximately 4.0, 4.2, 3.4, and 21.8 million parameters, respectively. These models were selected to represent a diverse range of lightweight and moderately complex architectures that are widely used in resource-constrained environments, such as CAVs, due to their balance between accuracy and computational efficiency. Training is performed 50 communication rounds with three clients, a batch size of 16, and 10 local epochs per client. We use stochastic gradient descent (SGD) as the optimizer, with a learning rate of 0.01, momentum of 0.9, and a weight decay factor of $4\times10^{-4}$. A StepLR scheduler is applied with a step size of 10 epochs and a gamma value of 0.1 to manage learning rate decay.

The experimental workflow proceeds through six key stages, as discussed. (1) Initialization and Key Generation: The central server defines the global model architecture (e.g., EfficientNet, MobileNet, ResNet) and configures encryption parameters, generating and distributing CKKS configurations using TenSEAL. (2) Local Training and Gradient Computation: Clients perform training on IID-simulated data and compute gradients, applying magnitude-based sensitivity maps to identify the most privacy-sensitive parameters. (3) Selective Encryption: Sensitive gradients are encrypted using CKKS, while non-sensitive ones remain in plaintext to minimize overhead. (4) Secure Aggregation: Clients send encrypted and plaintext gradients to the server, which performs homomorphic aggregation on ciphertexts and conventional aggregation on plaintext. (5) Decryption and Model Update: The server decrypts the aggregated ciphertext gradients and updates the global model with both encrypted and plaintext contributions. (6) Redistribution and Iteration: Updated model parameters are sent back to clients, and training continues iteratively until convergence. 

\section{Discussion}
This section presents the results and insights derived from applying the proposed CKKS-based selective homomorphic encryption scheme within a federated learning (FL) framework.

\subsection{Accuracy Comparison: With and Without Homomorphic Encryption}
\begin{figure}[H]
    \centering
    \includegraphics[width=1.0\linewidth]{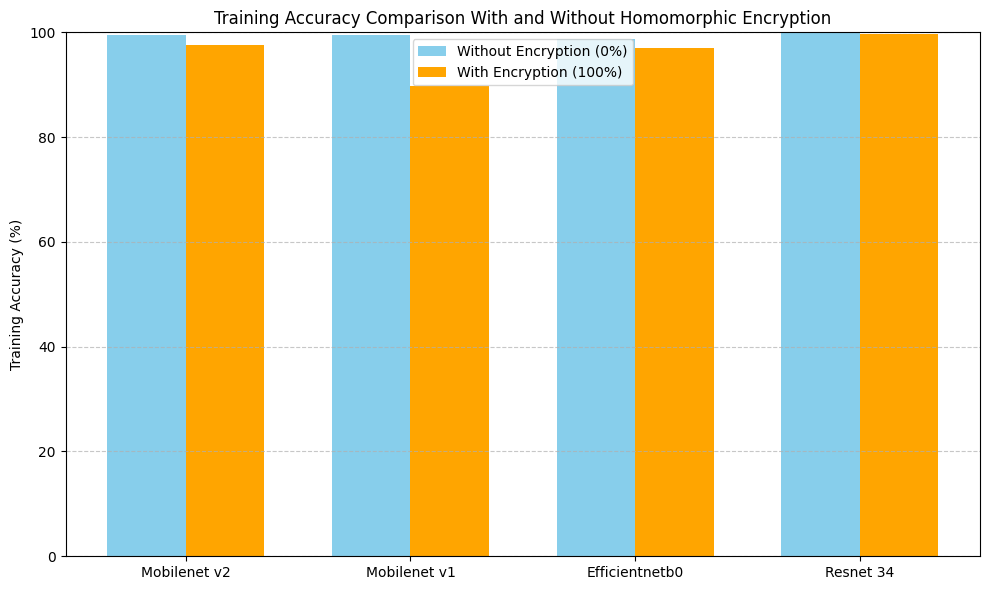}
    \caption{Comparison of training accuracy with and without homomorphic encryption, where 100\% encryption indicates that all model parameters are encrypted, and 0\% indicates no parameters are encrypted.}
    \label{training}
\end{figure}

\begin{figure}[H]
    \centering
    \includegraphics[width=1.0\linewidth]{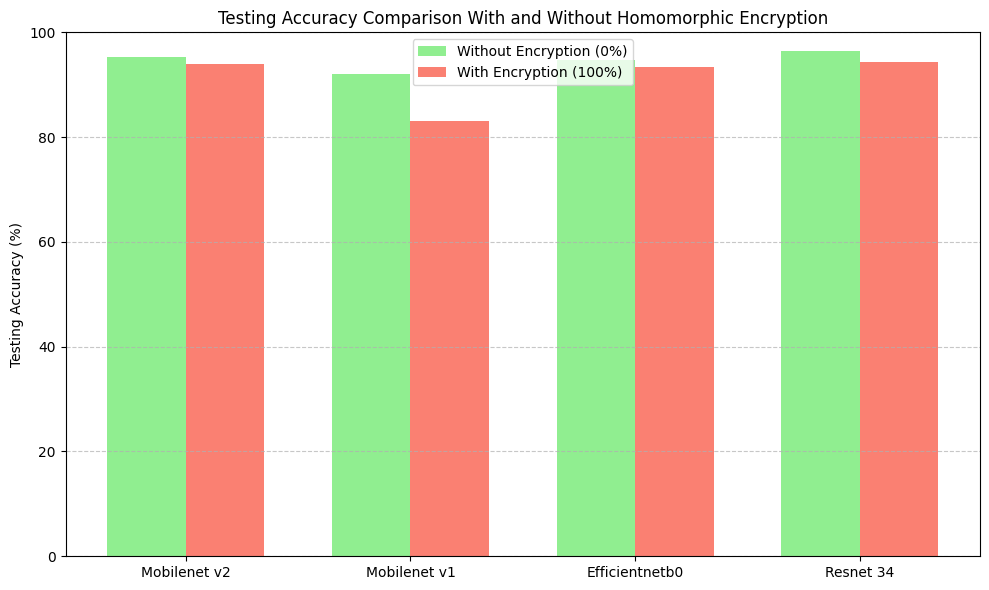}
    \caption{Comparison of testing accuracy versus model name with and without homomorphic encryption, where 100\% encryption indicates that all model parameters are encrypted, and 0\% indicates no parameters are encrypted.}
    \label{testing}
\end{figure}

To directly quantify the impact of HE on model accuracy, we present a comparative analysis for each model under two scenarios: training/testing \textbf{with} full encryption (100\%) and \textbf{without} encryption (0\%). Figures~\ref{training} and~\ref{testing} present the comparison of model accuracy with and without HE. 

Based on the results, ResNet34 and MobileNetV2 demonstrate strong robustness, maintaining high training and testing accuracy even under full encryption. EfficientNetB0 also shows relatively stable performance, with only a slight decline in accuracy. In contrast, MobileNetV1 experiences a noticeable drop, particularly in testing accuracy, under full encryption, indicating its greater sensitivity to the computational overhead introduced by homomorphic encryption.

Overall, these comparisons demonstrate that while homomorphic encryption introduces some measurable performance degradation, model accuracy remains largely preserved with and without encryption. Certain architectures are more affected due to inherent structural constraints, highlighting the importance of careful model selection when deploying fully encrypted training pipelines. This analysis emphasizes the need to balance privacy preservation with architectural suitability for encrypted federated learning.

\begin{figure}[htbp]
    \centering
    \includegraphics[width=1\linewidth]{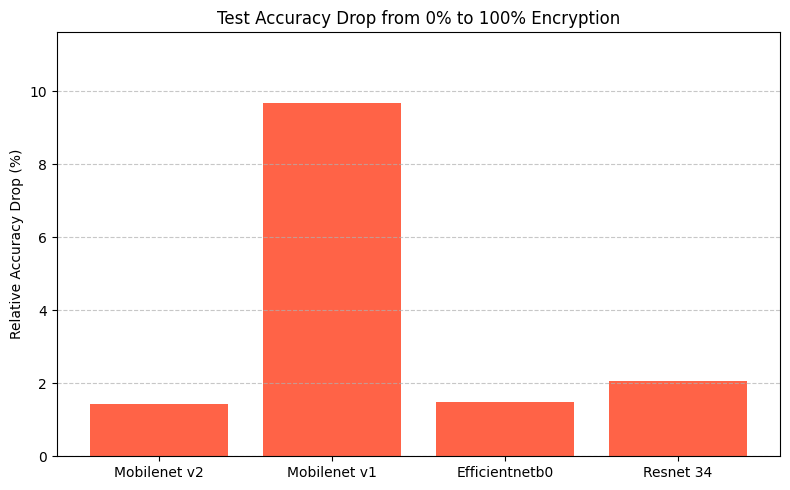}
    \caption{Relative drop in test accuracy (in percentage) for each model when moving from no encryption (0\%) to full homomorphic encryption (100\%).}
    \label{fig:rel_acc_drop}
\end{figure}

Next, following the lines of observing behavior in accuracy changes, we quantify the explicit impact of full encryption on predictive performance, Figure~\ref{fig:rel_acc_drop} displays the relative percentage decrease in test accuracy when moving from 0\% to 100\% encryption for each model.

\noindent

\subsection{Analysis on Generalization Gap}
The generalization gap, defined as the difference between training and test accuracy, serves as a critical indicator of model overfitting. By plotting the generalization gap for each model and encryption ratio as grouped bars, we reveal the sensitivity of each model to overfitting as privacy increases.

\begin{figure}[H]
    \centering
    \includegraphics[width=1.0\linewidth]{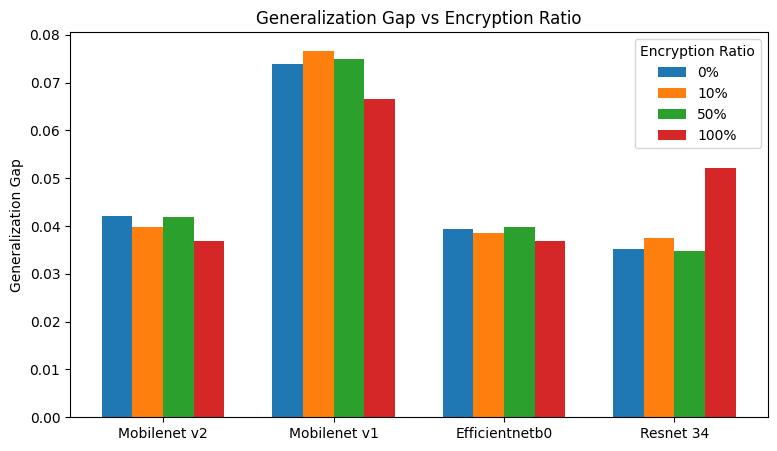}
    \caption{Generalization gap ($\mathrm{TrainAccuracy}_r - \mathrm{TestAccuracy}_r$) for each model at encryption ratios $0\%$, $10\%$, $50\%$, and $100\%$. Smaller values indicate better generalization.}
    \label{fig:gen_gap}
\end{figure}

As shown in Figure \ref{fig:gen_gap}, a lower bar indicates better generalization (less overfitting). The plot reveals how encryption affects generalization for different models, highlighting which architectures remain robust as privacy increases. As encryption ratios increase, the generalization gap typically widens, indicating that encryption can negatively affect generalization capabilities. Notably, MobileNetV1 experiences the most significant increase in the generalization gap at 100\% encryption, suggesting substantial vulnerability to overfitting under high encryption overhead. Conversely, ResNet34, EfficientnetB0, and MobileNetV2 maintain relatively smaller gaps even at higher encryption ratios, demonstrating robustness against generalization degradation caused by encryption overhead.

\subsection{Analysis of Model Size vs. Testing Accuracy}
\begin{figure}[htbp]
    \centering
    \includegraphics[width=1.0\linewidth]{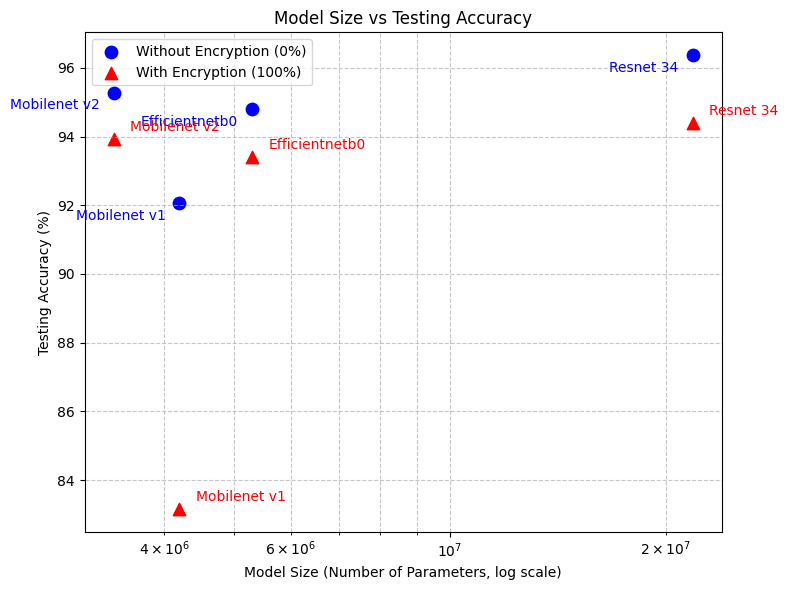}
    \caption{Testing accuracy as a function of model size (log scale) for both unencrypted (0\%) and fully encrypted (100\%) models.}
    \label{fig:size_acc}
\end{figure}

We investigate the relationship between model size (measured in the number of parameters) and testing accuracy under both unencrypted and fully homomorphically encrypted settings. Figure~\ref{fig:size_acc} plots each model’s parameter count on a logarithmic scale against its test accuracy at 0\% and 100\% encryption.

The results reveal several key trends. ResNet34 and EfficientNetB0, both with relatively high parameter counts, achieve the highest test accuracies in both encrypted and unencrypted settings, suggesting the advantages of increased model capacity. However, MobileNetV2 stands out by maintaining strong accuracy and robustness under encryption, despite being the smallest in size. This makes it a highly efficient option for resource-constrained or privacy-sensitive applications. In contrast, MobileNetV1 shows a notable drop in both test accuracy and privacy-preserving performance under full encryption, indicating that architectural design, not just model size, plays a critical role in resilience to homomorphic encryption overhead.

Overall, the transition to fully encrypted training leads to some accuracy degradation across all models, but the extent varies depending on architecture. While larger models can offer superior performance, they are not always the most practical under encryption constraints. EfficientNetB0 and MobileNetV2, in particular, strike a favorable trade-off between model complexity and encryption robustness. These findings underscore the importance of careful model selection in privacy-preserving federated learning, where both accuracy and computational efficiency must be balanced.

\subsection{Multi-metric Model Analysis for Accuracy and Efficiency}
To comprehensively compare the impact of encryption ratios on each model, we visualize four key metrics for each model and scenario using radar plots (see Figures~\ref{ratio_v2}, \ref{ratio_v1}, \ref{ratio_e0}, and \ref{ratio_res}) below. Each axis is constructed directly from the experimental data, without additional assumptions:

\begin{itemize}
    \item \textbf{Accuracy} ($A(r)$):\\
    Test set accuracy at encryption ratio $r$.
    \[
    A(r) = \mathrm{TestAccuracy}_r
    \]
    
    \item \textbf{Computational Efficiency} ($E_{\mathrm{comp}}(r)$):\\
    Normalized inverse training time, with $T_r$ denoting the training time (in hours) at ratio $r$:
    \[
    E_{\mathrm{comp}}(r) = 1 - \frac{T_r - T_{\min}}{T_{\max} - T_{\min}}
    \]
    where $T_{\min}$ and $T_{\max}$ are the minimum and maximum training times across all models and scenarios.
    
    \item \textbf{Generalization Efficiency} ($E_{\mathrm{gen}}(r)$):\\
    Normalized measure of the generalization gap between training and test accuracy:
    \[
    E_{\mathrm{gen}}(r) = 1 - \frac{(\mathrm{TrainAccuracy}_r - \mathrm{TestAccuracy}_r) - \Delta_{\min}}{\Delta_{\max} - \Delta_{\min}}
    \]
    where $\Delta_r = \mathrm{TrainAccuracy}_r - \mathrm{TestAccuracy}_r$, and $\Delta_{\min}, \Delta_{\max}$ are the minimum and maximum gaps across all experiments.
    
    \item \textbf{Efficiency of Training Loss} ($E_{\mathrm{loss}}(r)$):\\
    Normalized inverse of the average traning loss value:
    \[
    E_{\mathrm{loss}}(r) = 1 - \frac{\mathrm{Loss}_r - L_{\min}}{L_{\max} - L_{\min}}
    \]
    with $\mathrm{Loss}_r$ the average loss at ratio $r$, and $L_{\min}, L_{\max}$ the min/max values observed.
\end{itemize}

Radar plots effectively illustrate the trade-offs and performance stability across different models under varying encryption levels. MobileNetV2 consistently demonstrates strong computational efficiency and stable accuracy, making it particularly well-suited for scenarios requiring both performance and privacy. EfficientNetB0 achieves high accuracy but incurs greater computational overhead, indicating a trade-off between performance and resource consumption. ResNet34, while delivering excellent accuracy and low loss due to its large parameter count, suffers from significant computational penalties at higher encryption ratios. In contrast, MobileNetV1 shows notable vulnerability, with its accuracy degrading more rapidly as encryption intensity increases. These visual comparisons underscore the importance of balancing accuracy, efficiency, and encryption overhead when selecting models for privacy-preserving federated learning.

Overall, these plots enable an at-a-glance comparison of how privacy mechanisms affect each model, and support evidence-based model selection for privacy-preserving federated learning.

\begin{figure}[H]
    \centering
    \includegraphics[width=0.8\linewidth]{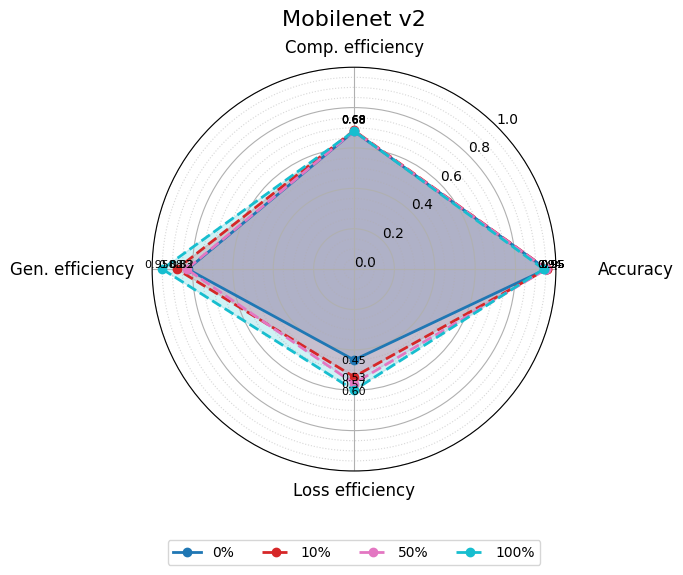}
    \caption{Performance trade-off analysis of MobileNetV2 under selective encryption.}
    \label{ratio_v2}
\end{figure}

\begin{figure}[H]
    \centering
    \includegraphics[width=0.8\linewidth]{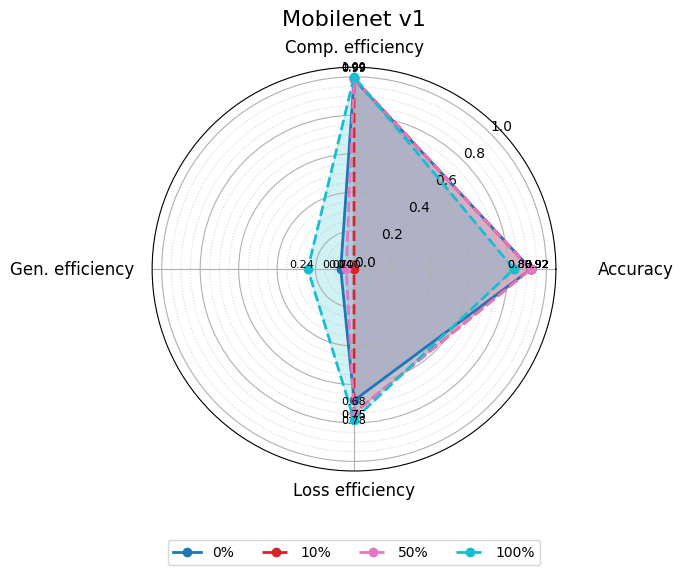}
    \caption{Performance trade-off analysis of MobileNetV1 under selective encryption.}
    \label{ratio_v1}
\end{figure}

\begin{figure}[H]
    \centering
    \includegraphics[width=0.8\linewidth]{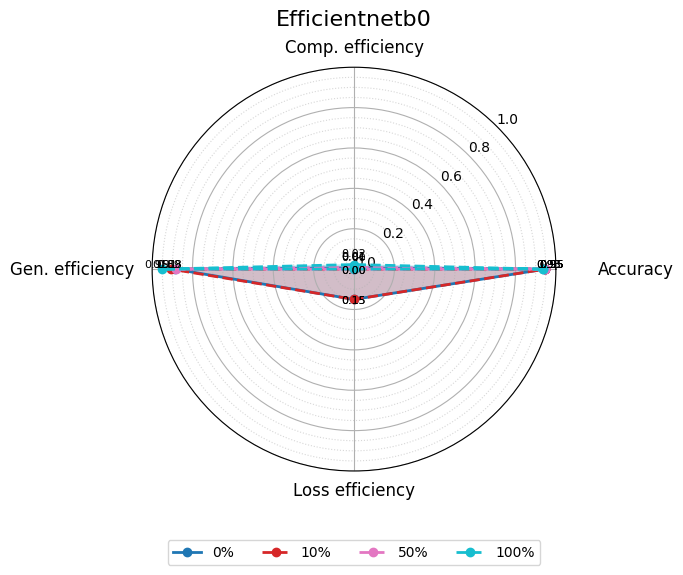}
    \caption{Performance trade-off analysis of EffecientnetB0 under selective encryption.}
    \label{ratio_e0}    
\end{figure}

\begin{figure}[H]
    \centering
    \includegraphics[width=0.8\linewidth]{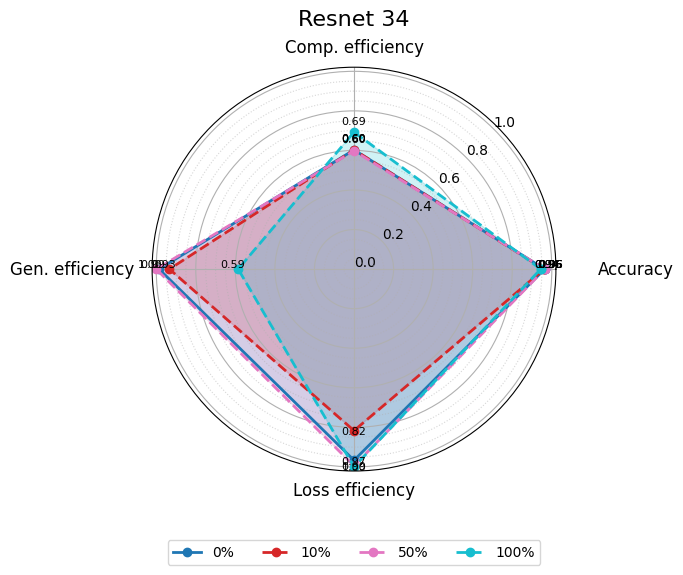}
    \caption{Performance trade-off analysis of Resnet34 under selective encryption.}
    \label{ratio_res}    
\end{figure}

In addition, the experimental results reveal a non-linear relationship between encryption ratio and model performance. Selective encryption effectively balances accuracy, computational cost, and privacy. As the proportion of encrypted parameters increases, all models exhibit the expected trade-offs, namely, slight reductions in computational efficiency and communication performance. However, these declines remain manageable, supporting the practicality of selective encryption for federated learning.

\section{Conclusion and Future Research Directions}
The integration of CKKS-based leveled homomorphic encryption using TenSEAL within a federated learning framework demonstrates strong potential for mitigating gradient leakage attacks while maintaining data privacy. Our experimental results on the CIFAR-10 dataset show that this approach achieves effective privacy preservation with minimal impact on model accuracy and comparable computational efficiency. These findings underscore the practical viability of selectively encrypted FL for real-world, privacy-sensitive applications—such as in connected and autonomous vehicles, and offer a compelling foundation for further research and deployment in secure collaborative machine learning systems.

While the findings of this research demonstrate the feasibility of integrating federated learning and homomorphic encryption for real-world applications such as connected and autonomous vehicles, several critical challenges remain. These include managing computational overhead, fine-tuning encryption parameters, and ensuring scalability. Future research should explore hardware acceleration techniques, as HE imposes significant computational costs; leveraging trusted execution environments (TEEs) and other specialized hardware could make large-scale encrypted FL more viable. Another promising direction is hybrid privacy-preserving approaches that combine HE with methods like Secure Multi-Party Computation or Differential Privacy to strike better trade-offs between security, efficiency, and model accuracy. Additionally, there is a need for dynamic encryption parameter adaptation, where encryption strength and resource usage can be adjusted in real time based on threat levels or model sensitivity. To address bandwidth and latency concerns, reducing communication overhead through techniques such as model sparsification, compression, or selective encryption remains an open challenge. Lastly, progress in this field would benefit from standardized benchmarking frameworks to evaluate HE-enabled FL systems across consistent metrics, including accuracy, training time, and privacy guarantees.

In addition, future research in CAV-focused FL should address several key challenges to enhance the practicality and security of privacy-preserving systems. First, low-latency encrypted model updates are essential to reduce communication delays and support real-time decision-making in vehicular environments. Second, heterogeneous hardware integration must enable efficient encrypted computation across the diverse and resource-constrained embedded platforms within CAVs. Third, robustness against adversarial threats remains critical, requiring advanced defenses not just for DLG but against poisoning, model inversion, and gradient leakage attacks. Finally, cooperative edge infrastructure, such as roadside units (RSUs) and fog nodes, should be leveraged to facilitate secure, encrypted model aggregation and distribution among high-mobility vehicles. Advancing these directions will help pave the way for secure, intelligent, and privacy-aware autonomous systems, contributing to the future of smart and trustworthy transportation networks. 

\section*{Acknowledgment}
This research was partially supported by Embry-Riddle Aeronautical University Faculty Innovative Research in Science and Technology (FIRST) grant.

\bibliographystyle{ieeetr} 
\bibliography{references}

\end{document}